\documentclass[twocolumn]{aastex61}
\usepackage{amsmath}
\usepackage{subfigure}
\input epsf
\usepackage{natbib}
\shorttitle{Progenitor of young Type Ia}
\shortauthors{Sarbadhicary et al.}

\begin{document}

\title{The two most recent thermonuclear supernovae in the Local Group: radio constraints on their progenitors and evolution}

\correspondingauthor{Sumit K. Sarbadhicary}
\email{sks67@pitt.edu}

\author{Sumit K. Sarbadhicary}
\affiliation{Pittsburgh Particle Physics, Astrophysics and Cosmology Center (PITT PACC), University of Pittsburgh, 3941
O'Hara St, Pittsburgh, PA 15260, USA}

\author{Laura Chomiuk}
\affiliation{Department of Physics and Astronomy, Michigan State University, East Lansing, MI 48824, USA}

\author{Carles Badenes}
\affiliation{Pittsburgh Particle Physics, Astrophysics and Cosmology Center (PITT PACC), University of Pittsburgh, 3941
O'Hara St, Pittsburgh, PA 15260, USA}

\author{Evangelia Tremou}
\affiliation{Department of Physics and Astronomy, Michigan State University, East Lansing, MI 48824, USA}

\author{Alicia M. Soderberg}
\affiliation{Harvard-Smithsonian Center for Astrophysics, 60 Garden St., Cambridge, MA 02138, USA}

\author{Lor\'{a}nt O. Sjouwerman}
\affiliation{National Radio Astronomy Observatory, P.O. Box O, 1003 Lopezville Rd., Socorro, NM, 87801}



\begin{abstract}
Young supernova remnants (SNRs) provide a unique perspective on supernova (SN) progenitors and connect the late evolution of SNe with the onset of the SNR phase. Here we study SN 1885A and G1.9+0.3, the most recent thermonuclear SNe in the Local Group (with ages $\sim 100$ years) with radio data, which provides a sensitive probe of the SN environment and energetics. We reduce and co-add 4-8 GHz observations from Karl G. Jansky Very Large Array (VLA) to produce the deepest radio image of the M31 central region (RMS noise of 1.3 $\mu$Jy/beam at 6.2 GHz). We detect some diffuse emission near SN 1885A at 2.6 $\sigma$, but the association with SN 1885A is uncertain because diffuse radio emission pervades the M31 central region. The VLA upper limit and HST measurements yield an ambient density, $n_0 < 0.04$ cm$^{-3}$ ($\pm$ 0.03 cm$^{-3}$ due to systematics) for SN 1885A, and kinetic energies, $E_k \sim (1.3-1.7) \times 10^{51}$ ergs for ejecta masses of $1-1.4$ M$_{\odot}$. For the same ejecta mass range, VLA observations of G1.9+0.3 yield $n_0 = 0.18$ cm$^{-3}$, and $E_k = (1-1.3) \times 10^{51}$ ergs. We argue that a sub-Chandrasekhar explosion model may be likely for SN 1885A, in agreement with X-ray studies, but in tension with models for the HST absorption spectra. The analysis of G1.9+0.3 is consistent with both Chandrasekhar and sub-Chandrasekhar SN Ia models, but rules out Type Iax explosions.
\end{abstract}
\keywords{acceleration of particles --- radio continuum: galaxies --- ISM: supernova remnants --- stars: supernovae: individual (SN 1885A, G1.9+0.3)}
\section{Introduction}
Young, ejecta-dominated SNRs provide a unique way to constrain the progenitors and evolution of SNe. The forward shock can probe the circum-stellar medium around these SNRs on pc-scales, corresponding to the pre-supernova mass-loss history on timescales of $10^3 - 10^5 $ years, which provides an independent way of discriminating progenitor scenarios \citep{Badenes2006, Badenes2007, Yamaguchi2014, Patnaude2015, Patnaude2017}. Additionally, these objects can trace how late-stage SNe, i.e. older than a few years, transition to the supernova remnant (SNR) phase \citep{Milisavljevic2017}. Radio emission from late-stage SNe usually declines \citep{Stockdale2001b, Eck2002, Weiler2002, Stockdale2006}, but young SNRs usually brighten as the forward shock encounters the bulk interstellar medium (ISM) \citep{Berezhko2004, Green2008}. However, this transition phase is very short compared to the lifetime of an SNR \citep{Sarbadhicary2017}, and therefore relatively unconstrained by observations. The oldest radio-detected SN is SN 1923A \citep{Eck1998} at an age of 80 years, while the youngest SNR (before the discovery of G1.9+0.3) was Cassiopeia A at 325 years, located in our Galaxy \citep{Fesen2006}. Fortunately, there are two objects in the Local Group that fill this gap, both with thermonuclear origins: SN 1885A in M31 and SNR G1.9+0.3 in our Galaxy.

SN 1885A, or S And, is a unique Local Group SN that exploded in 1885, about 65 pc from the center of M31. It was the first and one of the brightest recorded extragalactic SN, with $M_V = -19.2$ \citep{deVaucouleurs_Corwin85}. The expanding ejecta of SN 1885A was first imaged in optical \emph{absorption} in the Fe I resonance band against the backdrop of the M31 bulge \citep{Fesen1989}. Observations of its historical V-band light curve, color evolution and reconstructed spectrum pointed to a thermonuclear origin \citep{deVaucouleurs_Corwin85}. Subsequent observations by the \emph{HST Space Telescope} \citep{Fesen1999, Hamilton2000, Fesen2007, Fesen2015, Fesen2016} revealed stratified Ca and Fe layers, indicative of a Type Ia origin. The unusually fast light curve of SN 1885A, with a rise-time of 6 days and $\Delta m_{15} (B) \sim 2.2$ \citep{deVaucouleurs_Corwin85, Perets2011}, makes it part of an emerging class of fast-transients that are being discovered by high-cadence surveys \citep[See e.g.][]{Kasliwal2012, Kasliwal2012a, Foley2013, Drout2013, Drout2014, Taubenberger2017}. These transients extend the explosive endpoints of stellar evolution beyond the classic core-collapse and thermonuclear SNe. In this regard, SN 1885A provides a unique perspective to studying transients because we not only have observations of the historical light curve and spectra, but also of the freely expanding ejecta. However, SN 1885A has not shown any detectable emission in optical \citep{Fesen1989}, X-ray \citep{Williams2006, Li2009} or radio \citep[][, this work]{Crane1992}.

Here we present the deepest radio image at the site of SN 1885A by co-adding archival observations from the Karl G. Jansky Very Large Array (VLA), yielding an upper limit on the radio flux that is almost a factor of 3 deeper than  \cite{Crane1992}. We use this upper limit to constrain the radio light curve model from \cite{Sarbadhicary2017} model, exploring the parameter space of ambient density, and progenitor kinetic energy and ejecta mass for SN 1885A. We also apply this model to SNR G1.9+0.3, extending the constraints on its progenitor mass and kinetic energy from previous studies \citep{Reynolds2008, Ksenofontov2010, Carlton2011, deHorta2014, Yang2016, Chakraborti2016, Pavlovic2017}.

The paper is organized as follows - Section \ref{sec:obs} details the VLA observations and data reduction for the inner arc-minute of M31, Section \ref{sec:sn85radio} describes the final, co-added 6.2 GHz image, with particular focus at the site of SN 1885A, Section \ref{sec:model} summarizes the radio light curve model of \cite{Sarbadhicary2017} and how it is applied to the observations from the previous sections, Section \ref{sec:results} shows the results of our parameter space studies and Section \ref{sec:disc} discusses these results in the context of progenitor scenarios and further evolution of SN 1885A and G1.9+0.3.


\begin{figure*}[htp]
\centering
\includegraphics[width=\textwidth]{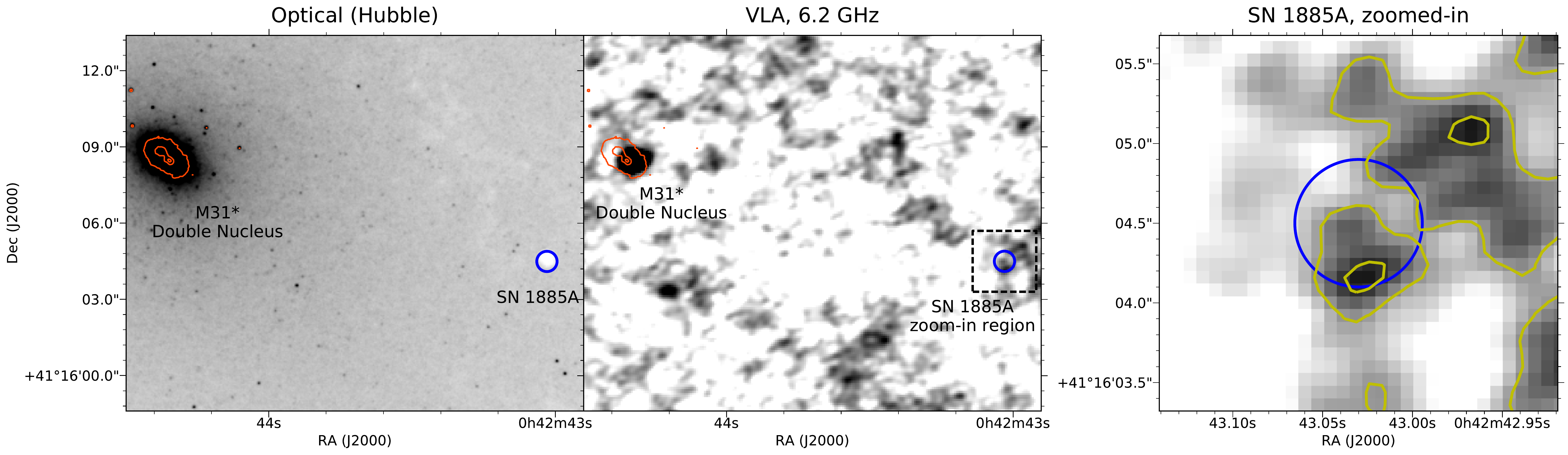}
\caption{Three-panel image of the vicinity of SN 1885A. \textit{Left}: HST FR388N image showing the \ion{Ca}{2} absorption (blue) in the SN 1885A remnant \citep{Fesen2007}. The red contours towards the top-left of the frame highlight the optical double-nucleus of M31, as imaged with \emph{HST}. The blue circle to the right outlines the \ion{Ca}{2} remnant of SN 1885A and has a radius of 0.4$^{\prime\prime}$. \textit{Center}: VLA 6.2 GHz radio continuum image, wth the optical contours (red) and the SN 1885A absorption profile overlaid as a blue circle. The dashed-square shows the region around SN 1885A zoomed-in in the next panel. \textit{Right}: Same VLA image, but zoomed in to emphasize the diffuse emission regions. The extent of the panel is the same as the dashed-square region from the central panel. The blue circle is also the same from the central panel, denoting the 0.4$^{\prime\prime}$ radius absorption profile of SN 1885A. The yellow contours trace the 6.2 GHz emission, with their lowest level at 1.82 $\mu$Jy beam$^{-1}$ (1.4 $\sigma$), and then at 3.64 $\mu$Jy beam$^{-1}$.}
\label{fig:sn1885im}
\end{figure*}

\begin{deluxetable*}{lccccccccc}
\tablewidth{0 pt}
\tablecaption{ \label{tab:obs}
Details of the archival VLA datasets for the central arc-minute of M31.}
\tablehead{ UT Date & Project Code & Configuration & Time On Source & Frequency & Bandwidth\\
 & & & (hr) & (GHz) & (MHz)}
\startdata
2011 May 19 & 11A-178 & BnA & 1.7 & 4.9 & 256 \\
2011 May 28  & 11A-178 & BnA & 1.9 & 4.9 & 256 \\
2011 Jun 5  & 11A-178 & BnA$\rightarrow$A & 1.9 & 4.9& 256 \\
2011 Jul 1  & 11A-178 & A & 1.9 & 4.9 & 256 \\
2011 Jul 27  & 11A-178 & A & 1.9 & 4.9 & 256 \\
2011 Aug 27 & 11A-178 & A & 1.9 & 4.9 & 256 \\
2011 Sep 2 & SD487 & A & 5.2 & 4.9 & 256 \\
2011 Sep 4 & 11A-137 & A & 0.7 & 6.8 & 2048 \\
2011 Sep 6 & 11A-137 & A & 0.7 & 6.8 & 2048 \\
2011 Sep 7 & SD487 & A & 5.2 & 4.9 & 256 \\
2012 Jun 1 & SD487 & B & 0.6 & 6.8 & 2048 \\
2012 Jun 6 & SD487 & B & 0.6 & 6.8 & 2048 \\
2012 Jun 12 & SD487 & B & 0.6 & 6.8 & 2048 \\
2012 Jul 1 & SD487 & B & 0.6 & 6.8 & 2048 \\
2012 Aug 14 & SD487 & B & 0.6 & 6.8 & 2048 \\
2012 Oct 29 & SD487 & A & 0.6 & 6.8 & 2048 \\
2012 Dec 22 & 12B-002 & A & 1.3 & 4.9/6.8 & 4096 \\
2012 Dec 30 & 12B-002 & A & 0.6 & 6.8 & 4096 \\
\enddata
\end{deluxetable*} 

\section{Radio Observations of the M31 core} \label{sec:obs}
We co-added 18 archival data sets obtained with the VLA in 2011 and 2012 (listed in Table \ref{tab:obs}) to produce a deep high-resolution image at 6.2 GHz. These data were observed by programs 11A-178 (PI Z.~Li), SD487 (PI M.~Garcia), 11A-137 (PI L.~Sjouwerman), and 12B-002  (PI L.~Sjouwerman), and were obtained to monitor M31's supermassive black hole, M31$^{\star}$ and the associated central region \citep[e.g.,][]{Garcia_etal10, Yang2017}. Observations were obtained in the A, B, and BnA configurations of the VLA, providing sub-arcsecond resolution and resolving out most of the diffuse emission from M31. All data were obtained in full Stokes mode, and because they were obtained over a period of Jansky VLA commissioning, bandwidths varied between 256--4096 MHz, and the central frequency also varied. For all wide-band data ($\geq$2048 MHz bandwidth), there were 14 spectral windows with frequency settings in common centered at 6.8 GHz; these are the windows used for imaging and they yield a total bandwidth of 1792 MHz. For the narrower band data (256 MHz of bandwidth), we used both spectral windows, which were centered at 4.9 GHz. Note that the 12B-002 epochs also included wide-band data centered at 4.9 GHz, and these data were ultimately averaged (in the image plane) with the narrower band data.

Data were calibrated using the complex gain calibrator J0038+4137 and the flux calibrator 3C48, using standard routines in AIPS \footnote{http://www.aips.nrao.edu/index.shtml}. Editing for bad antennae and radio frequency interference was carried out in AIPS using \verb|spflg| and \verb|tvflg|. We concatenated all wide-band data together and all narrower-band data together using the task \verb|dbcon| within AIPS. Imaging of each of the two concatenated data sets was performed using \verb|imagr| with natural weighting (Briggs Robust value of 5) and a minimum baseline length of 5 kilo-wavelengths. The lower frequency data (mostly narrower in bandwidth) yielded an image at 4.9 GHz with an rms sensitivity of 2.1 $\mu$Jy beam$^{-1}$ and a FWHM resolution of $0.67^{\prime\prime} \times 0.48^{\prime\prime}$. The wide band data produced an image at 6.8 GHz with an rms sensitivity of 1.6 $\mu$Jy beam$^{-1}$ and a FWHM resolution of $0.64^{\prime\prime} \times 0.60^{\prime\prime}$. To produce the deepest image possible, we smoothed both images to have a resolution of $0.68^{\prime\prime} \times 0.61^{\prime\prime}$ (synthesized beam position angle -80$^{\circ}$). We then co-added the narrow and wide band images, weighting by 1/(image noise)$^2$, yielding an image centered at 6.2 GHz with an rms noise of 1.3 $\mu$Jy beam$^{-1}$. As SN 1885A is offset from the image center (and the center of M31) by just 15.2$^{\prime\prime}$, a negligible fraction of the 3.3$^{\prime}$ radius primary beam, we do not apply a primary beam correction since it would be $ < 5 \%$.

\section{SN 1885A in the Radio} \label{sec:sn85radio}
To compare our radio images with archival narrow-band \emph{HST Space Telescope} images of SN 1885A, where the SN is seen in \ion{Ca}{2} H and K absorption (\citealt{Fesen2007}; also see the left panel of Figure \ref{fig:sn1885im}), we must first register the images to a common world coordinate system (WCS). The radio image is accurately placed on the J2000 system, as the phase calibrator has a positional accuracy $<$0.002$^{\prime\prime}$; therefore, uncertainties in the optical WCS dominate. 

The Local Group Galaxies Survey \citep[LGGS;][]{Massey_etal06} has imaged M31 in multiple optical filters and provides a WCS based on the USNO-B1.0 catalog, which has an rms astrometric accuracy of $\sim$0.2 arcsec \citep{Monet_etal03}. To place this on the International Celestial Reference System, we shifted the LGGS WCS by a small amount ($\Delta$ RA = -0.003s, $\Delta$ Dec = -0.13$^{\prime\prime}$) to come into optimal alignment with the UCAC2 and UCAC3 catalogs. The final astrometric accuracy of the LGGS image is 0.1$^{\prime\prime}$ rms (P. Massey, private communication), which is sufficiently accurate for this work.

The monochromatic field of view of the FR388N ramp filter on \emph{HST}/ACS is limited, and there are very few stars in common between the \ion{Ca}{2} image and the LGGS images. For this reason, we use an intermediary broad-band \emph{HST} image---an ACS F435W image observed in 2004 (HST program 10006; PI Garcia)---to provide a wider field of view and a solid comparison with the U-band LGGS Field 5 image. Using \emph{koords} of the KARMA software package \citep{Gooch96}, we register the images (allowing for both x/y shifts and rotation) based on six stars in common and find an rms of 0.02$^{\prime\prime}$. Next, we register the narrow-band \ion{Ca}{2} image to the newly-registered F435W image using a similar procedure; seven stars in common imply an rms of 0.0025$^{\prime\prime}$. Propagating through, the accuracy of the final WCS on the \ion{Ca}{2} image is 0.1 arcsec. 

The absorption-line remnant of SN 1885A is shown in the left panel in Figure \ref{fig:sn1885im}. It is centered at RA= 00$^{\rm{h}}$42$^{\rm{m}}$43.03$^{\rm{s}}$, Dec = +$41^{\circ}16^{\prime}04.5^{\prime\prime}$. As measured from its \ion{Ca}{2} absorption, the remnant has a radius of 0.40$^{\prime\prime}$ (1.5 pc at an assumed distance of 785 kpc), implying a time-averaged expansion velocity of $12,400\pm1,400$ km s$^{-1}$ \citep{Fesen2007}. The absorption profile would be just slightly resolved in our 6.2 GHz image. The central and right panels of Figure \ref{fig:sn1885im} show our radio maps at the position of SN 1885A, with a blue circle showing the position of the \ion{Ca}{2} absorption. Note that there is substantial sensitivity in our radio images on angular scales larger than the synthesized beam ($\sim$0.64$^{\prime\prime}$ diameter), as we co-added A and B configuration data together. 

Our naturally weighted image shows diffuse clumps of radio continuum emission pervading the inner regions of M31. This diffuse emission is present at a low level at the location of SN~1885A (peak flux 4.1 $\mu$Jy, or 2.6$\sigma$ significance), but the emission appears to be part of the background diffuse emission, and may not be distinct to the SNR. In other words, there are many other similar patches of emission present in M31's central region which are not coincident with a SNR, so there is no evidence that the low-level excess flux is attributable to SN 1885A.

To further investigate, we also imaged M31's central region using a Briggs Robust value of 0, which down-weights the shorter baselines and diffuse emission. This image has an rms sensitivity of 1.9 $\mu$Jy beam$^{-1}$ and a FWHM resolution of $0.34^{\prime\prime} \times 0.30^{\prime\prime}$. A 2.8 $\sigma$ peak in flux density is still visible coincident with the southwestern rim of SN 1885A (peak of 5.4 $\mu$Jy). Again, it is not clear if this flux is associated with SN~1885A or diffuse emission pervading the central regions of M31. We take this potential emission into account in calculating our upper limit on the flux of SN 1885A, adding the summed flux over a region to the 3$\sigma$ uncertainty borne of image noise. Assuming that the dimensions of SN 1885A match those of the \ion{Ca}{2} absorption, we find a 3$\sigma$ upper limit, $<$6.8 $\mu$Jy. 
If we instead allow for the possibility that the radio emission could be located at larger radius that the \ion{Ca}{2} absorption (coincident with the fastest moving material), and measure the radio upper limit in a region 0.60$^{\prime\prime}$ in radius (1.5 times the \ion{Ca}{2} SNR radius), we find a 3$\sigma$ upper limit, $<$11.4 $\mu$Jy. 
We take this conservative upper limit in our analysis throughout the rest of this paper.

As a side note, we also investigated the possible radio detection of SN 1885A mentioned by \cite{Sjouwerman_Dickel01}, offset by 1.3 arcsec from the center of the \ion{Ca}{2} remnant in our wcs-corrected image described above (claimed J2000 position of RA = 00$^{\rm{h}}$42$^{\rm{m}}$43.136$^{\rm{s}}$, Dec=$41^{\circ}16^{\prime}05.06^{\prime\prime}$.  \cite{Sjouwerman_Dickel01} find that this putative source has a flux, $27\pm10$ $\mu$Jy at 8.4 GHz, and is slightly extended in their images. We do not see any notable 6.2 GHz flux above 1$\sigma$ at this position, and  therefore believe this tentative detection of SN 1885A is erroneous.
\begin{figure*}
\subfigure[]{\includegraphics[width=\columnwidth]{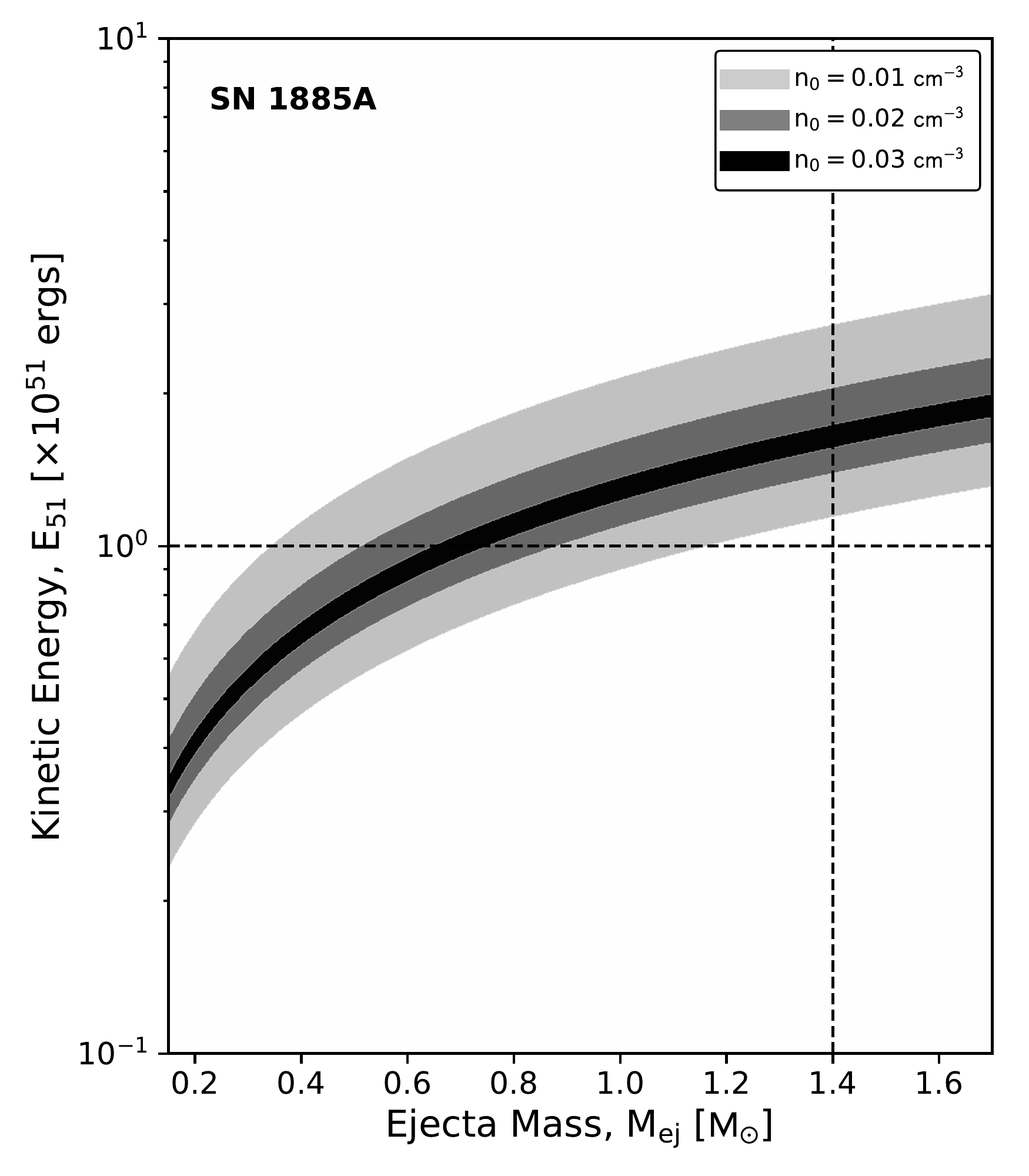}\label{fig:1885a}}
\subfigure[]{\includegraphics[width=\columnwidth]{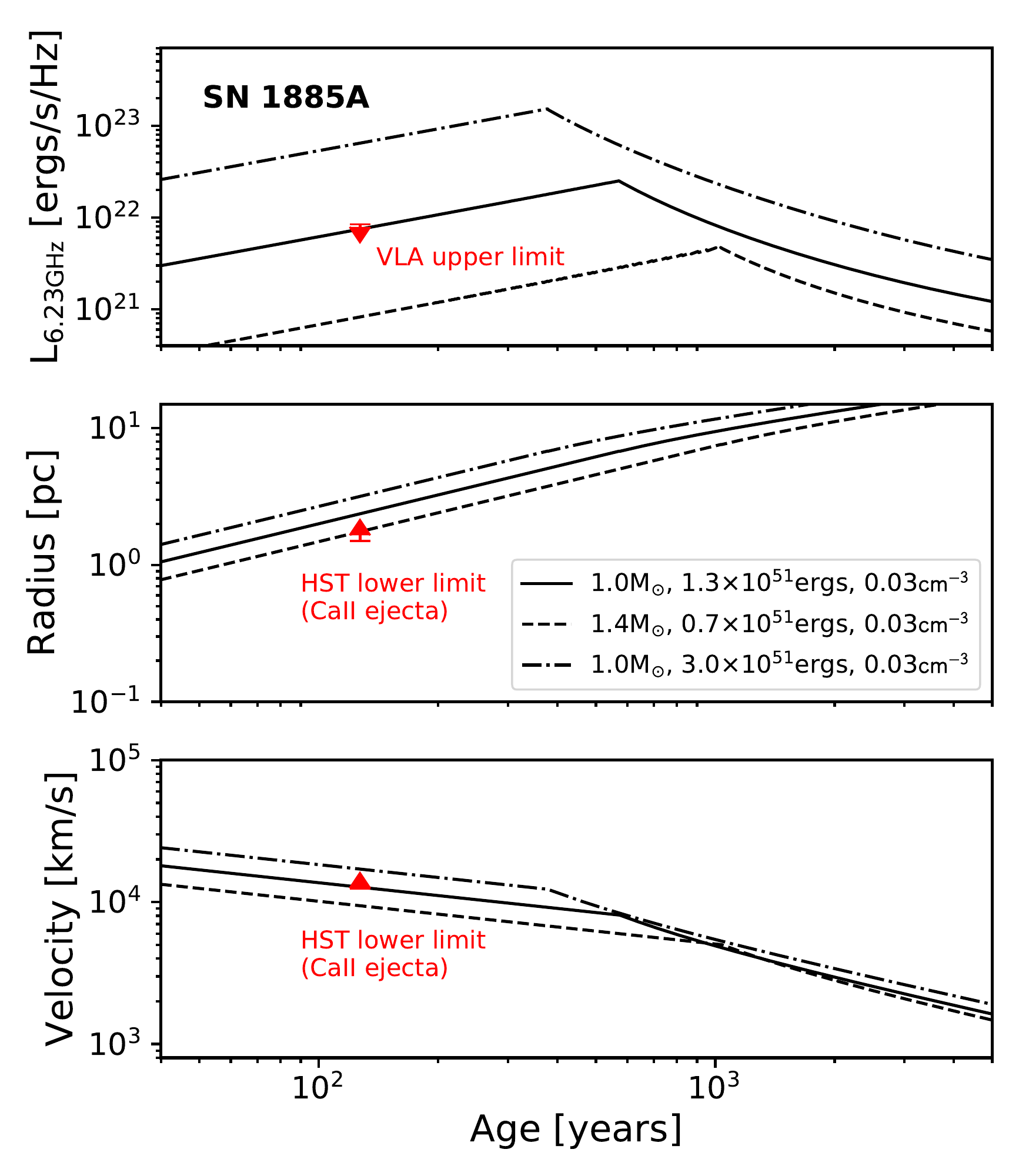}\label{fig:1885aLC}}
\caption{\textit{a)} Constraints on M$_{ej}$, E$_{51}$ and n$_0$ of SN 1885A from our upper limit on the radio flux from VLA, and the lower limits on shock radius and velocity from the outermost Ca-II ejecta from HST images. Each shaded band corresponds to a particular n$_0$, and shows all possible values of E$_{51}$ and M$_{ej}$ that simultaneously satisfy the three observational limits. The bands get narrower for higher n$_0$, as explained in Section \ref{sec:results}. The crosshairs mark the fiducial explosion parameters for SNe Ia: M$_{ej}$ = 1.4 M$_{\odot}$ and E$_{51}$ = 1. \textit{b)} Light curves, radii and velocity curves of SN 1885A for different values n$_0$, E$_{51}$ and M$_{ej}$. The solid line was chosen from darkest band in a) and is consistent with all three observational limits. The dashed and dash-dotted lines were chosen from outside the shaded parameter bands in a), and are not consistent with all the observational limits.}
\end{figure*}
\begin{figure*}
\subfigure[]{\includegraphics[width=\columnwidth]{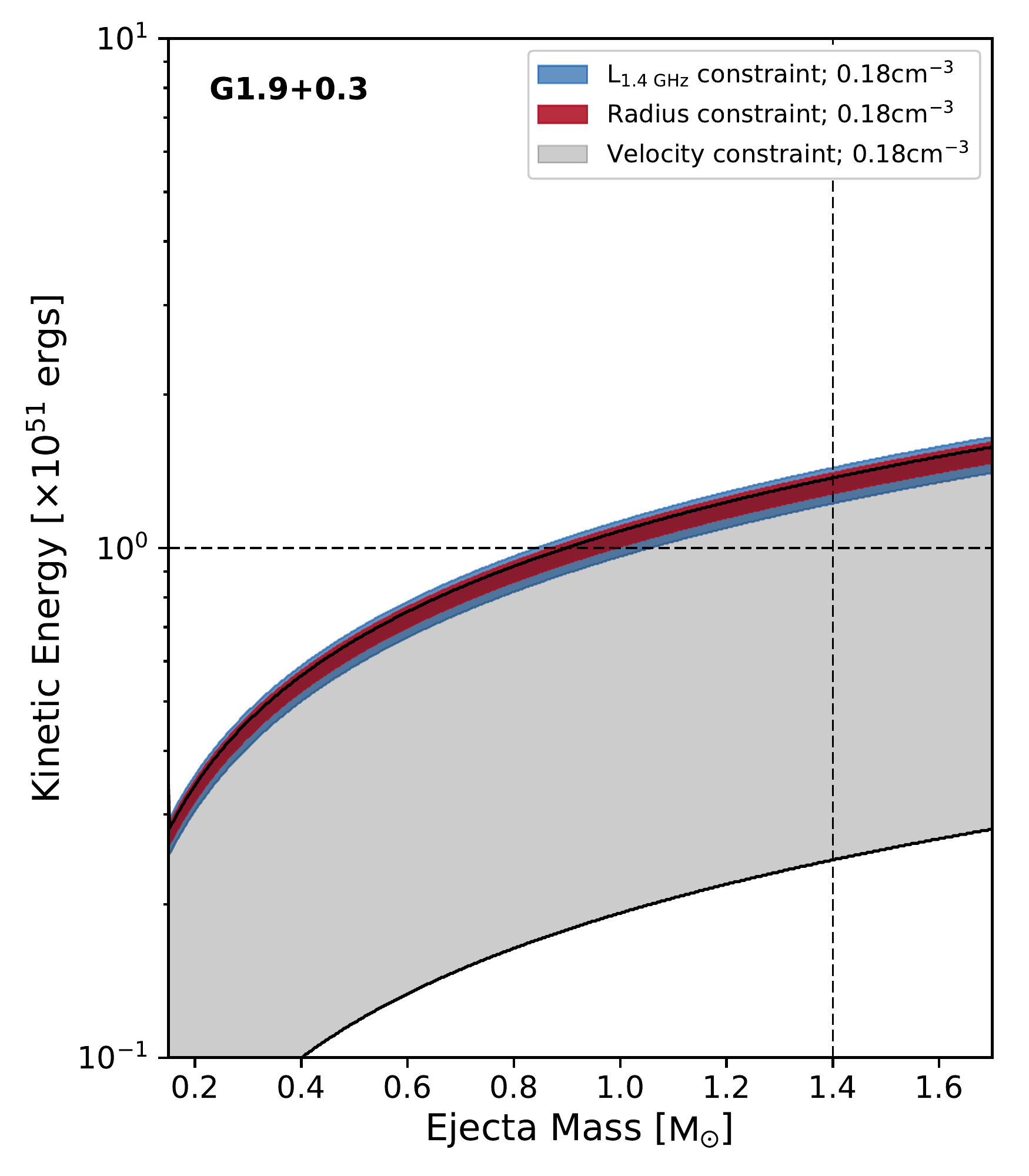}\label{fig:g19}}
\subfigure[]{\includegraphics[width=\columnwidth]{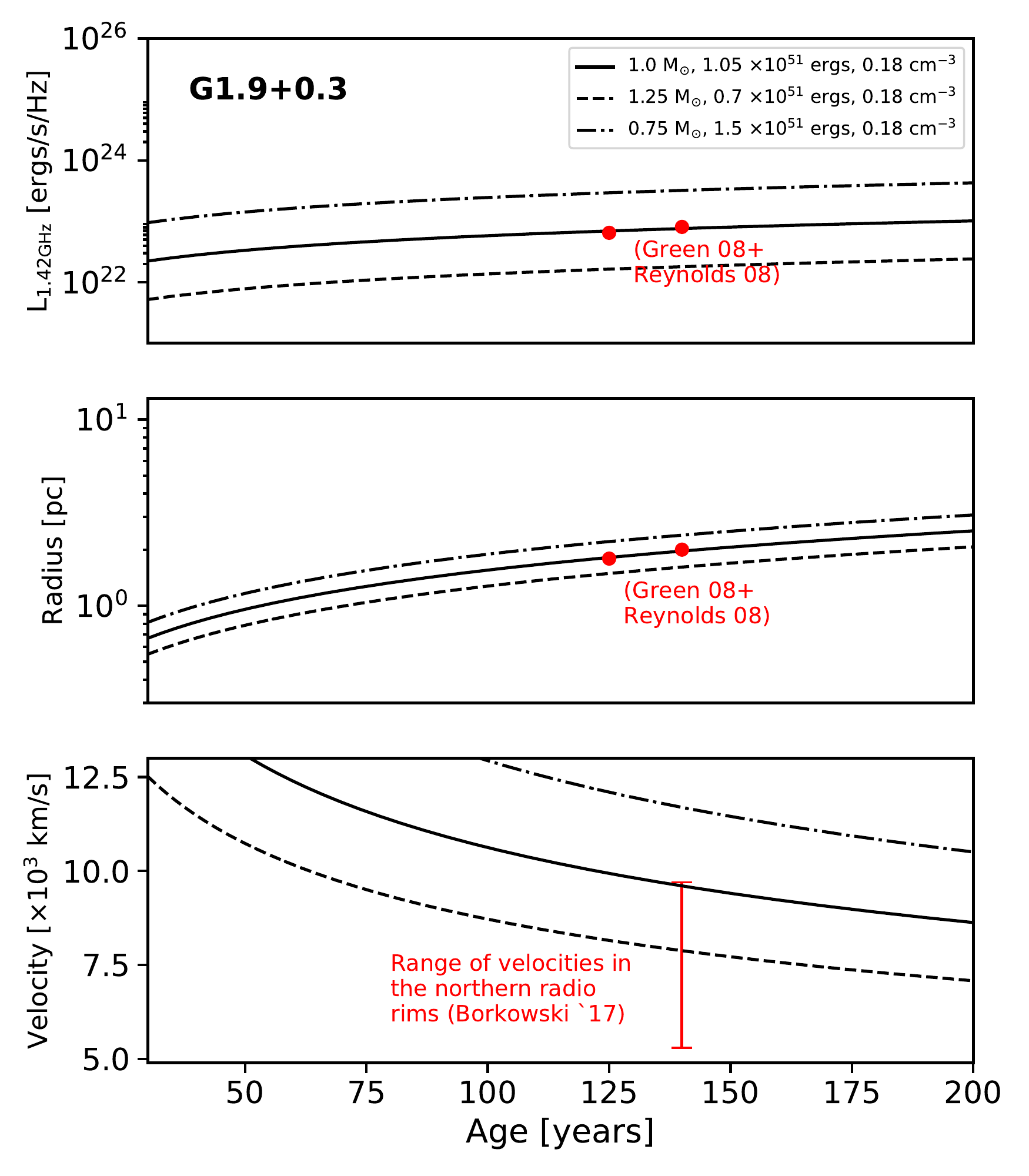}\label{fig:g19LC}}
\caption{Constraints on M$_{ej}$, E$_{51}$ and n$_0$ for G1.9+0.3 from archival VLA data. a) The blue and red band corresponds to values of M$_{ej}$ and E$_{51}$, for a given $n_0$, for which the model radio luminosities and radii, respectively, agree with observations in 1993 and 2008, as recorded in \cite{Green2008}. The grey band shows all parameter values for which the model shock velocities lie within the observed range in the northern radio rim. b) Light curves, radii and velocity curves for G1.9+0.3 produced for different values $n_0, E_{51}, M_{ej}$. The solid curves, which satisfy all three constraints, correspond to $n_0$, E$_{51}$ and M$_{ej}$ taken from common overlapping region of radio luminosity, radius and velocity parameter space in a), while the dashed and dash-dotted lines are taken from non-overlapping regions, and therefore do not satisfy all the observations.}
\end{figure*}

\section{Radio Light Curve Modeling}  \label{sec:model}
\subsection{Summary of model}
We will use the SNR radio light curve model described in \cite{Sarbadhicary2017} to constrain the ambient density ($n_0$, in units of cm$^{-3}$), kinetic energy ($E_k$), and ejecta mass ($M_{ej}$, in units of M$_{\odot}$) of SN 1885A and G1.9+0.3. While we refer the reader to the paper for details of the radio light curve physics \citep[see Appendix A in][]{Sarbadhicary2017}, we will briefly describe its basic features here. The radio emission is assumed to be optically-thin synchrotron emission which, according to diffusive shock acceleration theory, is produced by non-thermal electrons accelerated by the forward shock \citep{Axford1977, Krymskii1977, Bell1978, Blandford1978} as it interacts with the circum-stellar medium \citep{Chevalier1982, Chevalier1998, Berezhkho2004}. The forward shock radius ($R_s$) and velocity ($v_s$) is calculated self-similarly from \cite{Truelove1999}, 
\begin{subequations} \label{eq:rsvs}
\begin{align}
R_s &= \left(1.29 \rm{pc}\right) t_{100}^{0.7} n_0^{-0.1} E_{51}^{0.35} M_{ej}^{-0.25}, \\
v_s &= \left(8797 \rm{km\ s^{-1}}\right) t_{100}^{-0.3} n_0^{-0.1} E_{51}^{0.35} M_{ej}^{-0.25}, 
\end{align}
\end{subequations} 
assuming a power-law ejecta, $n=10$ \citep{Matzner1999, Chomiuk2012, Chomiuk2016}. Here $t_{100}$ is the SNR age in units of 100 years, and $E_{51} = E_k/10^{51}\ \rm{ergs}$. The ejecta dominated solution is valid until $t \lesssim \left(203.5\ \rm{yrs}\right) E_{51}^{-1/2} M_{ej}^{5/6} n_0^{-1/3}$.

The synchrotron luminosity $L_{\nu}$ is a function of the synchrotron-emitting volume, the strength of the magnetic field and the population of accelerated electrons, all of which depend on $R_s$ and $v_s$. A large magnetic field strength (few 100 $\mu$G, almost 2 orders of magnitude above the ambient field) is required for efficient acceleration of particles in SNR shocks \citep{Volk2005, Warren2005, Morlino2010}, which, in turn, helps produce more synchrotron emission per accelerated electron. In our model, the energy density in this amplified, upstream field is a fraction $\epsilon^u_b$ of the shock energy density ($ \propto n_0 v_s^2$). Motivated by semi-analytic studies \citep{Bell1978, Bell2004, Amato2009} and recent first-principle kinetic simulations \citep{Caprioli2014a, Caprioli2014b}, we assume $\epsilon^u_b = 1/2 \left(v_s/c + 1/M_A\right)$, where $M_A$ is the Alfv\'{e}n Mach Number. The first and second terms are contributions from non-resonant and resonant instabilities induced in the upstream magnetic field by the streaming particles. The energy density in the accelerated electrons is similarly assumed to be a fraction $\epsilon_e$ of the shock energy density, and characterized by a power-law spectra with index, $p$. However, $\epsilon_e$ is an uncertain parameter in our model because of the incomplete understanding of electron injection in non-relativistic shocks \citep{Park2015, Spitkovsky2016}. Typically, $\epsilon_e \sim 10^{-4}$ in young SNRs based on their multi-wavelength spectra \citep{Morlino2012} while $\epsilon_e \gtrsim 10^{-3}$ for a population of SNRs, where older, Sedov SNRs will dominate \citep{Sarbadhicary2017}. Meanwhile, the spectral index is usually $p=2.2 - 2.5$ for young SNRs \citep{Caprioli2012}. For both SN 1885A and G1.9+0.3, we assume $\epsilon_e = 10^{-4}$ and $p=2.2$, and then discuss the effects of changing these parameters in Section \ref{sec:results}.

\subsection{Application to SN 1885A} \label{sec:SN 1885A}
The model described above allows us to explore in detail the full parameter space of $n_0$, $E_{51}$ and $M_{ej}$ relevant to SN 1885A. Specifically, we look for values of $n_0$, $E_{51}$ and $M_{ej}$ that simultaneously satisfy the following observational limits at an age of 127 years,
\begin{enumerate}
\item \textit{Radio luminosity}: $L_{\nu} \leq 8.43 \times 10^{21}$ ergs s$^{-1}$ Hz$^{-1}$ at 6.2 GHz, which is the 3$\sigma$ upper limit of 11.4 $\mu$Jy derived in Section \ref{sec:sn85radio}, taking into account the patch of diffuse emission coincident with the SN. We assume a distance = 785 $\pm$ 25 kpc to M31 \citep{McConnachie2005}.
\item \textit{Radius}: $R_{s} \geq 1.52$ pc, which is the extent of outermost CaII emission in the \cite{Fesen2015} image. Since the forward shock wave cannot be behind the ejecta, we treat this as the lower limit to $R_s$.
\item \textit{Velocity}: $v_s \geq 12500$ km s$^{-1}$, which is the outer velocity component of the CaII absorption profile \citep{Fesen2015}, and is treated as a lower limit to $v_s$ for the same reason as above.
\end{enumerate}

\subsection{Application to SNR G1.9+0.3} \label{subsec:g19}
The detectable emission in radio from G1.9+0.3 over the last 2 decades can provide stronger constraints on the parameter space of $n_0, E_{51}$ and $M_{ej}$ than SN 1885A \citep[see][and references therein]{Green2008, deHorta2014}. The objective is to find all values of $n_0, E_{51}$ and $M_{ej}$ for which the model predicts luminosities, radii and velocities that satisfy all the following observations within the uncertainties of each measurement,
\begin{enumerate}
\item \textit{Radio luminosity}: A 1.4 GHz integrated flux of 0.74 $\pm$ 0.038 Jy in 1993 \citep{Condon1998} and 0.935 $\pm$ 0.047 Jy in 2008 \citep{Green2008} was measured for G1.9+0.3. At a distance of 8.5 kpc \citep{Reynolds2008}, these fluxes correspond to luminosities of $\left(6.4 \pm 0.3 \right) \times 10^{22}$ ergs s$^{-1}$ Hz$^{-1}$ and $\left(8.1 \pm 0.4\right) \times 10^{22}$ ergs s$^{-1}$ Hz$^{-1}$ respectively.
\item \textit{Radius}: The mean radius of G1.9+0.3 in 2008 was 2 pc, based on the X-ray profile of \cite{Reynolds2008}, and therefore 1.79 pc at the time of the \cite{Condon1998} measurement, if we assume a constant expansion rate of 0.65 $\%$ per year \citep{Green2008}.
\item \textit{Velocity}: A shock velocity range of 5300 km s$^{-1}$ - 9700 km s$^{-1}$ is measured in the bright northern radio rim of G1.9+0.3, reflecting the significant asymmetry in the expansion, where 5300 km s$^{-1}$ is the median velocity in the N-NE region, and 9700 km s$^{-1}$ is the median velocity in the NW region \citep{Borkowski2017}. Because of this wide range of observed velocities, we first constrain the parameter space allowed by the range of velocities, then see where the constraints from the luminosity and radius measurements overlap.
\end{enumerate}

We also fix the following parameters in our model based on multi-wavelength observations, 
\begin{enumerate}
\item The ages of G1.9+0.3 at the time of \cite{Reynolds2008} measurement was 140 years, and \cite{Condon1998} was 125 years.
\item $\epsilon_e = 10^{-4}$, which is consistent with the measured electron-to-proton ratio from the non-thermal radio and X-ray spectra \citep{Ksenofontov2010, Pavlovic2017} and upper limits on cosmic-ray acceleration efficiency from gamma-ray observations \citep{HESS2014}. 
\item $p=2.2$, which means the synchrotron spectral index = $(p-1)/2 = 0.6$, which is the same as measured from the integrated radio emission in \cite{Green2008}.
\end{enumerate}

\section{Results} \label{sec:results}
\subsection{SN 1885A} \label{sec:results:1885a}
Figure \ref{fig:1885a} shows the constraints on the parameter space of SN 1885A from our analysis in the previous section. The shaded regions show values of $n_0, E_{51}, M_{ej}$ that predict luminosities, radii and shock velocities that are consistent with the observational limits in Section \ref{sec:SN 1885A}. We see that a range of M$_{ej}$ and E$_{51}$ are plausible, and higher values of E$_{51}$ are associated with higher M$_{ej}$. This is because higher M$_{ej}$ leads to smaller energy per unit mass of the forward-moving ejecta, and therefore a higher kinetic energy is required to produce the same radio luminosity, shock radius and velocity. As we increase n$_0$, the parameter space in Figure \ref{fig:1885a} shrinks. This is because for higher densities, a) SNRs are brighter as more energy is available to the shocked ambient medium, and therefore to the population of accelerated electrons and magnetic fields, and b) SNRs decelerate faster because of more resistance from the denser ambient medium (see Eq. \ref{eq:rsvs}). As a result, a smaller range of E$_{51}$ is allowed by the observational limits for a given M$_{ej}$. At n$_0$ = 0.04 cm$^{-3}$, the range of parameter space allowed by the three observational limits vanishes, and we can treat this as an upper limit on the ambient density. For densities just below this upper limit, we find that E$_{51}$ = 0.9-1.8 for M$_{ej}$ = 0.6-1.5 M$_{\odot}$, the typical range of ejecta masses inferred for SNe Ia \citep{Scalzo2014}, and E$_{51}$ = 1.3-1.7 for M$_{ej}$ = 1-1.4 M$_{\odot}$.  Although SN 1885A is not yet producing detectable radio emission, $E_k > 10^{51}$ ergs is still required by our model to explain the highest velocity component of the CaII ejecta. In Section \ref{sec:disc}, we will further discuss our measured density upper limit and the M$_{ej}$-E$_{51}$ parameter space in the context of SN Ia progenitor models and observations of the M31 central region.

To check the validity of the parameter space, we show the predicted radio light curves, radii and shock velocities for different values of $n_0, E_{51}$ and $M_{ej}$ in Figure \ref{fig:1885aLC}. For example, n$_0$ = 0.03 cm$^{-3}$, $M_{ej} = 1.2$ M$_{\odot}$ and $E_{51} = 1.4$ fall in the darkest region in Figure \ref{fig:1885a}, and predict luminosities below the radio upper limit, velocities and radii above the optical lower limits.  On the other hand, any set of values outside the shaded regions either predict luminosities above the upper limit, or velocities and radii below the lower limit. Note from Figure \ref{fig:1885aLC} that the shock velocity provides a stronger lower limit to our parameter space than radius. 
\begin{figure*} 
\centering
\hspace{0.1in}
\includegraphics[width=0.8\textwidth]{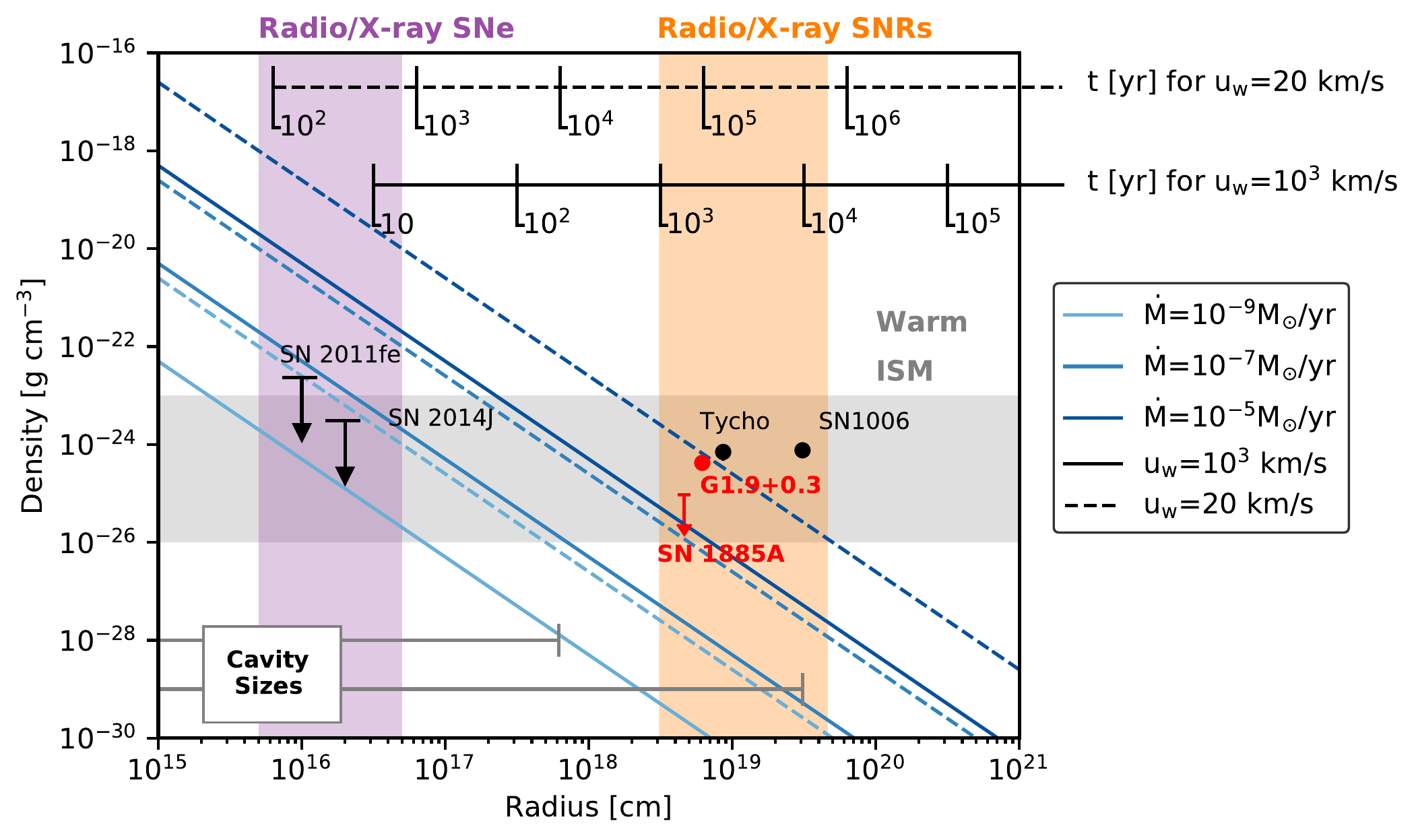}
\caption{Circum-stellar densities and radii of well-known Type Ia SNe and confirmed Type Ia SNRs studied in radio or X-rays, along with SN 1885A and G1.9+0.3 from this work. The dashed and dotted lines indicate different CSM profiles created by isotropic mass loss at constant rates and constant velocities given in the legend box. The ruler on top shows the timescales on which these CSM profiles formed in the pre-explosion stage. The grey region shows the typical extent of the warm ISM phase \citep{Ferriere2001}. The horizontal grey bars show the cavity sizes created by accretion-driven outflows from the progenitor due to fast and slow winds \citep{Badenes2007}. \textit{Reproduced from \cite{Patnaude2017}}\\
\textbf{Object references:} SN 2011fe \citep{Chomiuk2012}, SN 2014J \citep{Perez2014}, Tycho \citep{Slane2014}, SN 1006 \citep{Raymond2007}}
\label{fig:rhorad}
\end{figure*}

We also checked the robustness of our results to changes in our assumptions about electron acceleration, particularly the parameters $\epsilon_e$ and $p$. We recall that the upper limit on n$_0 = 0.04$ cm$^{-3}$ was for $\epsilon_e = 10^{-4}$ and $p=2.2$. For $\epsilon_e = 10^{-3}$ and $p=2.2$, the upper limit reduces to $n_0 \sim 0.01$ cm$^{-3}$, and predicted energies are lower than Figure \ref{fig:1885a}. This is because efficient electron acceleration makes SNRs brighter, and the model requires even lower ambient densities and energies to remain consistent with the dim radio upper limit. For $\epsilon_e = 10^{-4}$ and $p=2.5$, we get a higher density upper limit of $n_0 = 0.07$ cm$^{-3}$. This is because a steeper energy spectrum produces fewer electrons at GeV-energies that contribute to the synchrotron luminosity in the GHz range. As a result, the model can still satisfy the radio upper limit with larger values of n$_0$. Taking these systematics into account, $n_0 = 0.07$ cm$^{-3}$ is the highest possible value of the upper limit in our model, given the observational limits in Section \ref{sec:SN 1885A}. Deep, multi-wavelength observations in the future will help put stronger, independent constraints on the population of accelerated particles in SN 1885A.

\subsection{G1.9+0.3} \label{sec:results:g19}
Figure \ref{fig:g19} shows the M$_{ej}$-E$_{51}$ parameter space of G1.9+0.3 that are consistent with the observational constraints in Figure \ref{fig:g19LC}. The parameter space constrained by velocity is the largest because of the wide range of velocities seen in the northern radio rim. Since SNRs are brighter and smaller for higher n$_0$, the parameter spaces constrained by radio luminosity and radius will change in opposite directions for varying n$_0$, which allows a measurement of $n_0$ from the common parameter space of radio luminosity, radius and velocity. We find that G1.9 evolved in $n_0 \approx 0.18$ $\rm{cm^{-3}}$ and exploded with E$_{51}$ = 0.8-1.4 for M$_{ej}$ = 0.6-1.5 M$_{\odot}$, or E$_{51}$ = 1-1.3 for M$_{ej}$ = 1-1.4 M$_{\odot}$. However, lower $E_{51}$ and $M_{ej}$ are not ruled out by our parameter space. 

As with SN 1885A, we checked the validity of parameter space by showing the predicted light curves, shock radii and velocities for different parameter values in Figure \ref{fig:g19LC}. Values of n$_0$, E$_{51}$ and M$_{ej}$ that fall in the common area of parameter space constrained by radio luminosity, radius and velocity in Figure \ref{fig:g19} are consistent with our observations, whereas any value taken from outside the common area are not consistent with all three constraints.

We re-checked out results for different ages of G1.9+0.3, given the uncertainty in the kinematically-derived age measurements \citep{Reynolds2008, Carlton2011, deHorta2014}. We find that our model cannot simultaneously satisfy all the observational limits - luminosity, radius and the velocity range - for ages $<$ 140 years. For ages $>$ 140 years, we can satisfy the observations for n$_0 \gtrsim$ 0.1 cm$^{-3}$. It is likely therefore that G1.9 was a slightly less energetic explosion, but evolving in a denser ambient medium compared to SN 1885A. In Section \ref{sec:disc:g19}, we discuss these constraints further in the context of SN Ia models and previous measurements in the literature.
\begin{figure*}
\subfigure[]{\includegraphics[width=\columnwidth]{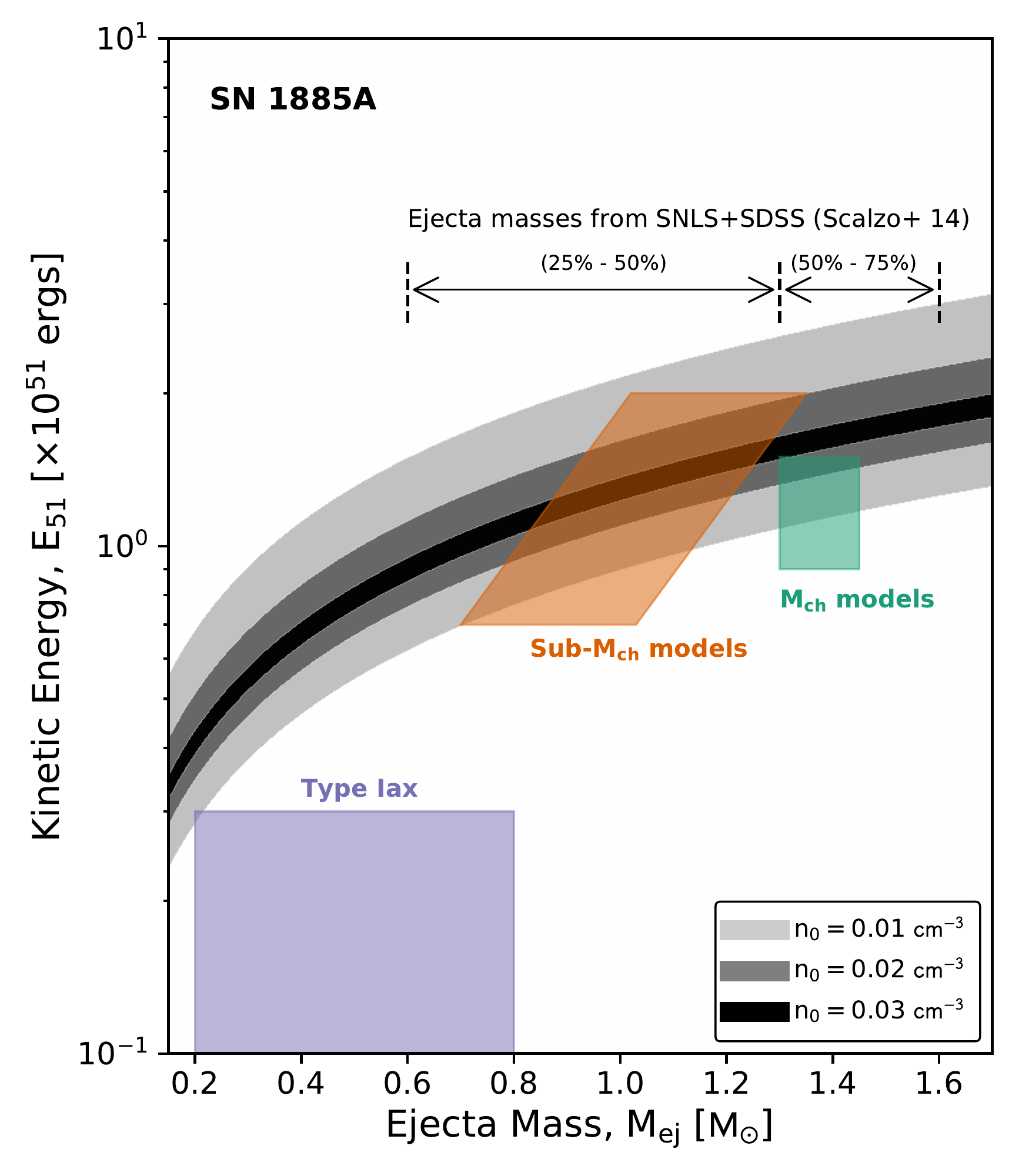} \label{fig:1885models}}
\subfigure[]{\includegraphics[width=\columnwidth]{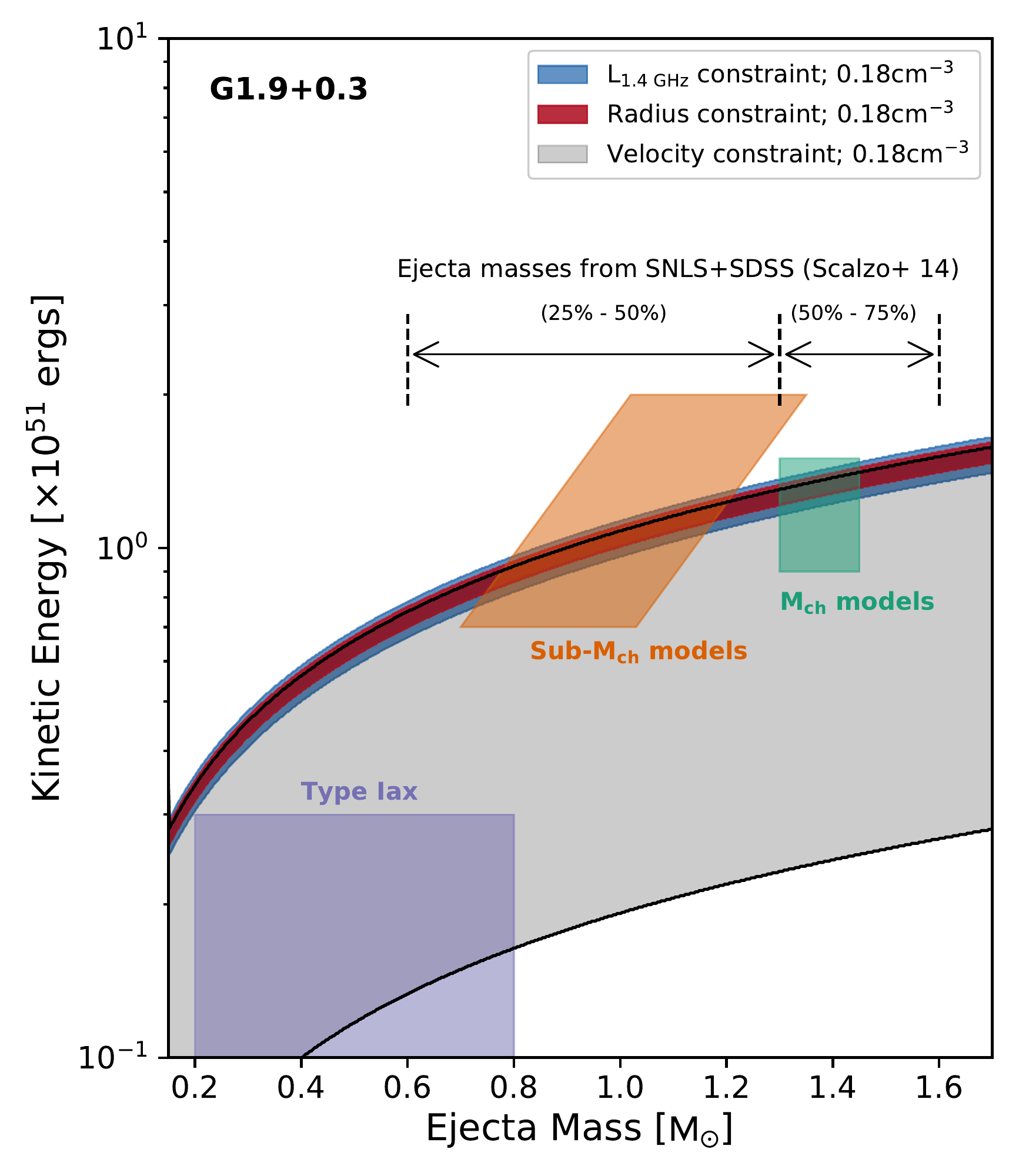} \label{fig:g19models}}
\caption{Same as Figures \ref{fig:1885a} and \ref{fig:g19}, but overlaid with different SN Ia progenitor models. The parameter space of M$_{ch}$ models is taken from \cite{Iwamoto1999, Badenes2003}, Sub-M$_{ch}$ models from \cite{Woosley2011, Shen2017} and Type Iax from \cite{Foley2013, Foley2016}. The arrows are meant to indicate the typical range of M$_{ej}$ for SNe Ia, and are taken from the distribution of M$_{ej}$ measured by \cite{Scalzo2014} from a sample of 337 cosmological SN Ia from SDSS and SNLS.}
\end{figure*}
\section{Discussion} \label{sec:disc}
\subsection{Progenitor scenarios of SN 1885A}
\subsubsection{Clues from the density upper limit}
The radio non-detection of SN 1885A in our work indicates that it is probably not encountering ambient medium with densities $> 0.07$ cm$^{-3}$. A low-density ambient medium was also inferred by \cite{Perets2011} from the \emph{Chandra} X-ray upper limit \citep{Li2009}. This measurement can shed some light on plausible progenitor scenarios for SN 1885A. In the standard single-degenerate (SD) scenario, a white dwarf accretes mass from a non-degenerate companion and explodes once it gets close enough to the Chandrasekhar limit, M$_{ch}$ = 1.4 M$_{\odot}$ \citep{Whelan1973, Li1997, Tutukov1996, Han2004}. Because the accretion process is likely non-conservative, some kind of circum-stellar medium (CSM) is expected around SD SN Ia progenitors \citep[see][for reviews]{Wang2012, Maoz2014}. If the material lost is ejected from the surface of the WD, large outflow velocities (several 1000 km s$^{-1}$) are expected, and the CSM should take the form of a large, low-density, energy-supported cavity. For lower outflow speeds, most of the material might remain close to the progenitor in a $\rho \propto r^{-2}$ structure \citep[see Figure 4 in][for a discussion and examples of CSM structures]{Badenes2007}. Our current deep radio limit, as well as past non-detections of SN 1885A in radio, makes any progenitor scenario that produces dense CSM close to the progenitor unlikely \citep{Pooley1967, Spencer1973, Dickel1984, Crane1992}. This is also the case for those Type Ia SNe that have not been detected in the radio \citep{Panagia2006, Hancock2011, Chomiuk2016} or X-rays \citep{Russell2012}, including the closest ones, SN2011fe \citep{Chomiuk2012, Margutti2012}  and SN2014J \citep{Perez2014, Margutti2014} - see Figure \ref{fig:rhorad}. At larger radii, the shock dynamics and ionization properties of most known Type Ia SNRs are inconsistent with evolution inside low-density cavities carved out by high velocity progenitor outflows \citep{Badenes2007, Yamaguchi2014, Patnaude2017}.

It is likely therefore that SN 1885A is simply expanding inside the bulk ISM, similar to other known Type Ia SNe and SNRs, instead of a CSM modified by the progenitor (Figure \ref{fig:rhorad}). In fact, SN 1885A is located inside the M31* circumnuclear region, i.e. the innermost few hundred parsecs, which is mostly dominated by warm-ionized and diffuse X-ray emitting hot gas, with no significant detection of the denser neutral or molecular phases \citep{Braun2009, Li2009, Li2011}. The average density in this region is about 0.1 cm$^{-3}$ \citep{Dosaj2002, Garcia2005},  which, along with our measured upper limit, is consistent with densities in the warm ISM phase (Figure \ref{fig:rhorad}). This may favor WD-WD progenitor models, which are not as strictly associated with accretion-driven mass loss and presence of CSM, unlike the SD models described previously \citep[although see][]{Shen2013, Raskin2013}. The merger timescales of such WDs \citep{Maoz2012} are also comparable to the stellar age distribution near SN 1885A \citep{Saglia2010}.

\subsubsection{Clues from kinetic energy-ejecta mass constraints}
The constrained M$_{ej}$-E$_{51}$ parameter space for SN 1885A can be compared with values that are usually associated with Chandrasekhar (M$_{ch}$) and sub-Chandrasekhar (sub-M$_{ch}$) models for normal Type Ia SNe (Figure \ref{fig:1885models}, model references are in the caption). In M$_{ch}$ models, the accreting WD mass has to be near M$_{ch}$ to ignite C in the degenerate core \citep{Iwamoto1999, Maeda2010, Blondin2013}. Sub-M$_{ch}$ models provide an alternate route to SN Ia, characterized by M$_{ej}$ $<$ M$_{ch}$, through a \emph{double-detonation}, i.e. detonation of an accreted layer of He on the surface of the primary WD, leading to a second detonation in the degenerate C/O core \citep{Nomoto1982a, Woosley1994, Bildsten2007, Woosley2011, Shen2017}. One can see that for densities within the upper limit, the constrained parameter space coincides with a larger area of the M$_{ej}$-E$_{51}$ parameter space for sub-M$_{ch}$ models than for M$_{ch}$ models. The combined constraints on ambient density in the previous section, and the M$_{ej}$-E$_{51}$ parameter space therefore appear to disfavor a M$_{ch}$ origin for SN 1885A. This is consistent with \cite{Perets2011}, which also showed that M$_{ch}$ delayed-detonation models cannot reproduce the \emph{Chandra} X-ray upper limit. In fact, explosions $<$ M$_{ch}$ can be expected in almost 25-50$\%$ of SNe Ia in the local universe \citep{Scalzo2014}. Our inference from radio is at odds, however, with the distribution of Ca and Fe in the \emph{HST} absorption profile of the ejecta, which favors an off-centered, M$_{ch}$ delayed-detonation explosion. \citep{Fesen2007, Fesen2015}.

Motivated by the fast light curve of SN 1885A, we also checked if our results are consistent with the properties of Type Iax SNe - a class of transients that are similar to Type Ia, but characterized by low peak luminosities, ejecta velocities and masses \citep[see][for references and review]{Jha2017}. \cite{Chevalier1988} and \cite{Perets2011} estimated M$_{ej}$ $\sim 0.1-0.3$ M$_{\odot}$ and $E_k \sim 2.2 \times 10^{50}$ ergs from the historical light curve of SN 1885A, and these estimates are similar to that of known Type Iax SNe \citep{Foley2013, McCully2014}. In Figure \ref{fig:1885models}, we show a plausible parameter space for Type Iax SNe with the purple box, derived from estimates of ejecta masses and expansion velocities given in \cite{Foley2013, McCully2014, Foley2016} and \cite{Jha2017}. The range of $E_k$ for SN 1885A, based on our limits on radio flux, optical radius and velocity, is still in general higher than that of Type Iax SNe. On the other hand, \cite{Garcia2017} recently showed that 3D models of interacting C/O or O/Ne WDs that are invoked for Ca-rich transients \citep{Perets2010, Kasliwal2012} can produce Ca and Fe distribution that looks similar to SN 1885A \citep{Fesen2015}. Thus, while our observational constraints do not energetically favor a Type Iax origin for SN 1885A, we cannot rule out the possibility that SN 1885A originated from some class of thermonuclear transient that is different from SNe Ia.

\subsection{Progenitor scenarios for G1.9+0.3} \label{sec:disc:g19}
Our constraints on the parameter space for G1.9+0.3, although tighter than SN 1885A, is less constraining on the common SN Ia progenitor scenarios. Our measured n$_0$ = $0.18$ cm$^{-3}$ from the radio emission of G1.9+0.3 is higher than the estimated n$_0 = 0.02$ cm$^{-3}$ obtained by modeling the shock dynamics of G1.9+0.3 inferred from X-ray images \citep{Reynolds2008, Carlton2011, Pavlovic2017}, but in agreement with analysis of VLA-ATCA images \citep{deHorta2014}, which infers n$_0 \lesssim 0.3$ cm$^{-3}$ using the numerical model of \cite{Berezhko2004}. Nevertheless, these densities are still much higher than densities expected inside wind-blown cavities, and within the density range of the warm ISM phase (see Figure \ref{fig:rhorad}). A dense, inhomogeneous ambient medium around G1.9+0.3 may not be surprising given the large difference in expansion rates around the remnant \citep{Borkowski2017} and the brightening radio emission. It is unclear though, if this ambient medium is associated with the progenitor, or if G1.9+0.3 is encountering an over-dense region of the ISM. The allowed range of E$_{51}$ inferred in Section \ref{sec:results:g19} can be satisfied by both M$_{ch}$ and sub-M$_{ch}$ models, but not by the lower-energy range of Type Iax SNe (Figure \ref{fig:g19models}). These results are at least consistent with detailed analyses of the X-ray abundances and ionization properties of the ejecta of G1.9+0.3 \citep{Borkowski2010, Borkowski2013, Yamaguchi2014} which suggested an asymmetric, delayed-detonation explosion. We also note that unlike SN 1885A, we have no constraints from the optical light curve, which would have been undetectable because of the high line-of-sight absorption towards G1.9+0.3 \citep{Reynolds2008}.
\begin{figure}
\includegraphics[width=\columnwidth]{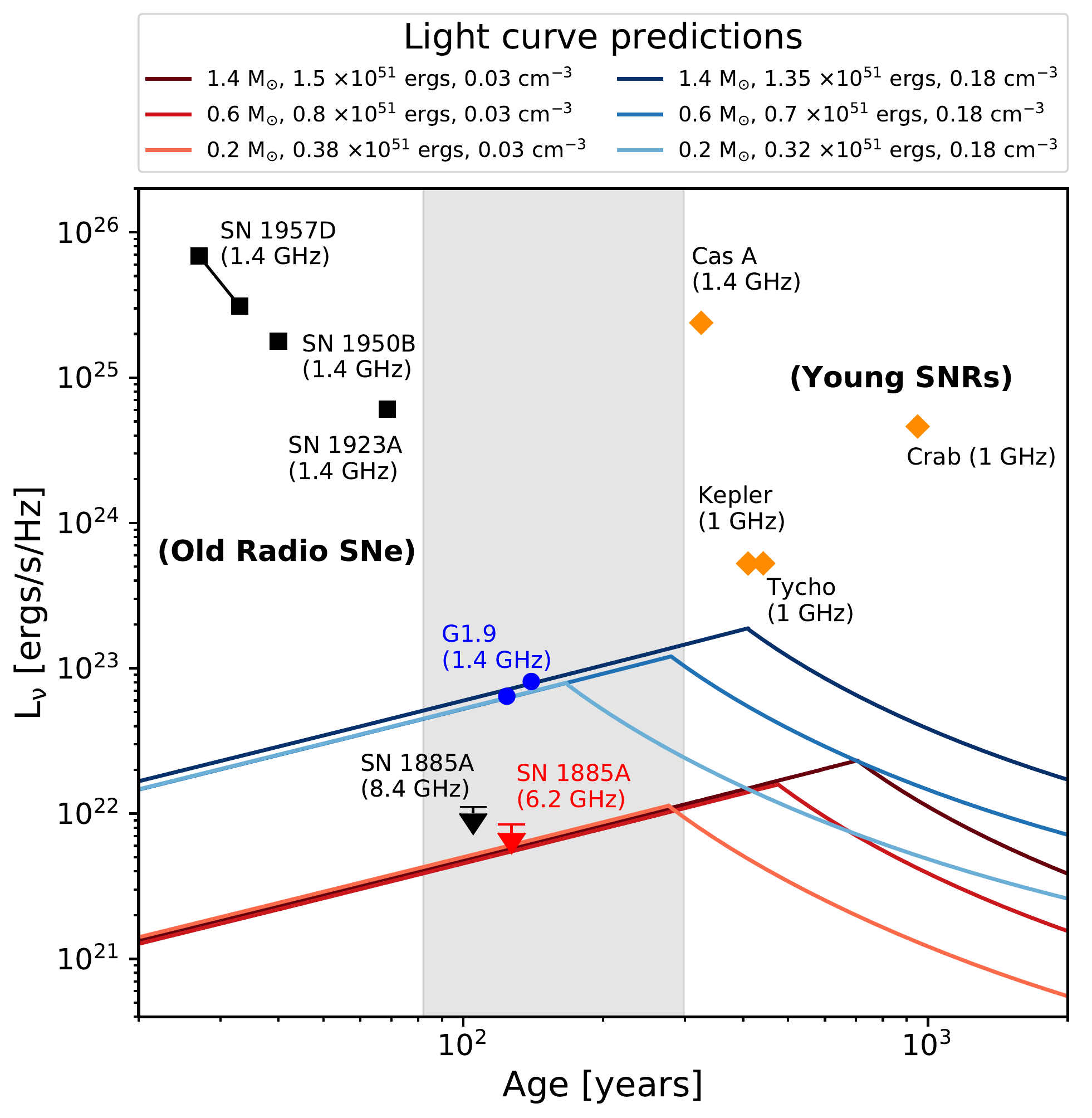}
\caption{Light curve predictions of SN 1885A and G1.9+0.3 from values of energy, M$_{ej}$ and ambient densities constrained from radio and optical absorption images. The blue points show the two 1.4 GHz measurements of G1.9+0.3 from \cite{Condon1998} and \cite{Green2008}. The 8.4 GHz (black) and 6.2 GHz (red) upper limits on SN 1885A are from \cite{Crane1992} and this work, respectively. Also shown are some of the oldest SNe and the youngest SNRs, from both thermonuclear and core-collapse origin, with detectable radio emission. The gray shaded region is where our sample is lacking in SNe and SNRs, and objects like G1.9+0.3 and SN 1885A hold the key to constraining this period of evolution.\\
\textbf{Object references:} SN 1957D, 1950B \citep{Cowan1994}, SN 1923A \citep{Eck1998}, Cas A, Kepler, Tycho and Crab \citep{Fesen2006, Green2014, Yamaguchi2014, Green2017, Trotter2017}}
 \label{fig:maxbright} 
\end{figure}

\subsection{The Supernova to Supernova-Remnant transition}
We can also make predictions about the future radio evolution of SN 1885A and G1.9+0.3 from their known ages, and constrained properties from Section \ref{sec:results}. While we are not certain if SN 1885A is producing any radio emission, the upper limit is still consistent with a young ejecta-dominated SNR expanding in a very low density ambient medium and is expected to brighten over time. Assuming that SN 1885A is just below the detection limit of 11.4 $\mu$Jy (Figure \ref{fig:maxbright}), we predict that it will brighten to a light curve maximum of about 15 $\mu$Jy in 6.2 GHz images in roughly another 150 years, if it was a very low energy explosion, or $\sim 31 \mu$Jy in another 570 years if it was a normal, energetic Type Ia. If SN 1885A went off in lower densities, the brightening period will be even longer. 

For the similarly aged G1.9+0.3, radio brightening is already evident. In fact, at the location of SN 1885A in M31, G1.9+0.3 would have had a radio brightness = (0.93 Jy) $(0.0085\  \mathrm{Mpc})^2/(0.785\ \mathrm{Mpc})^2 = 109\ \mu$Jy (or $\sim 45$ $\mu$Jy at 6.2 GHz), which would have been detectable at a level of almost 28$\sigma$. This means that SNe can be transitioning to the SNR phase at $\sim 100$ years, if there is enough CSM present for producing radio emission. 

We expect G1.9+0.3 to brighten for roughly another 270 years up to a flux of about 2.2 Jy in 1.4 GHz, if we assume a normal, Type Ia explosion. The brightening period is smaller than the prediction of \cite{Pavlovic2017} mostly because we measure almost an order of magnitude higher ambient density around G1.9+0.3 as a result of keeping the M$_{ej}$ - E$_{51}$ parameter space flexible in Section \ref{sec:results}.

It is interesting to note that the age and radio emission from both SN 1885A and G1.9+0.3 in Figure \ref{fig:maxbright} fall in the range where the radio `turn-on' is expected for SNRs \citep{Cowsik1984}. Future high resolution and high sensitivity radio interferometers may find more SN 1885A-like objects, which will help observationally constrain this transition region.

\section{Conclusions}
We presented the deepest radio image at the site of SN 1885A by reducing and co-adding $\sim$29 hours of 4-8 GHz VLA data of the M31 central region between 2011-2012. The final 6.2 GHz image has an RMS noise sensitivity of 1.3 $\mu$Jy, and shows regions of diffuse emission at 2.6$\sigma$ near SN 1885A. This emission is still present even if the site is re-imaged by down-weighting the shorter baselines. It is inconclusive whether the emission is uniquely associated with SN 1885A because such diffuse emission patches are scattered throughout the M31 central region. We assume a conservative 3$\sigma$ upper limit of 11.4 $\mu$Jy on the radio flux throughout the paper, taking into account this diffuse emission, and the possibility that the forward shock extends beyond the optical absorption region of SN 1885A. 

We explored the parameter space of ISM density-kinetic energy-ejecta mass of SN 1885A and G1.9+0.3 using the radio light curve model of \cite{Sarbadhicary2017}. Based on the upper limit on the radio flux from VLA, and the lower limit on the shock radius and velocity from HST, SN 1885A may have exploded in a very low density environment with an upper limit of 0.04 cm$^{-3}$ ($\pm 0.03$ cm$^{-3}$, if systematic uncertainties from the electron acceleration model is taken into account). For a density of 0.03 cm$^{-3}$, SN 1885A would have exploded with kinetic energies of $(1.3-1.7) \times 10^{51}$ ergs for ejecta masses in the range of $1-1.4$ M$_{\odot}$, usually associated with SN Ia. For even lower ambient densities, the range of kinetic energies allowed by observations is larger. The same light curve model was applied to the similarly aged SNR G1.9+0.3 which already had measurements of the integrated flux, shock radii and velocities from VLA and \textit{Chandra} observations. These observations constrained an ambient density of 0.18 cm$^{-3}$ around G1.9+0.3, denser than the ambient medium around SN 1885A, and kinetic energies of $(1 - 1.3) \times 10^{51}$ ergs for the same range of ejecta masses.

These measurements provided some clues about the progenitors of these objects. While G1.9+0.3 seems consistent with both M$_{ch}$ and sub-M$_{ch}$ explosion models, SN 1885A seems less likely to have originated from a M$_{ch}$ explosion, based on the energetics, the very low ambient density, and the lack of any radio detection in the past. This conclusion is supported by the observation and analysis of \emph{Chandra} X-ray upper limit by \citep{Perets2011}, but is at odds with the analysis of the ejecta from optical absorption, and the primordial light curve. A Type Iax origin is also energetically disfavored for both SN 1885A and G1.9+0.3.

The measured flux and known ages of G1.9+0.3 and SN 1885A make them ideal candidates for studying the `turn-on' phase of radio emission in SNRs. Our combined VLA and optical constraints suggest that SN 1885A will brighten for at least another 150 years (if it was a very low energy explosion) or up to 570 years, for a normal, 1.4 M$_{\odot}$ explosion. For G1.9+0.3, the brightening period is expected to continue for another $\sim 270$ years, assuming a normal 1.4 M$_{\odot}$ explosion with energy $> 10^{51}$ ergs. 

These results also underscore the importance of deep multi-wavelength observations and monitoring of SN 1885A in the future. While the observational limits in this study provided some hints about the origin and evolution of SN 1885A, an actual detection of the ejecta will resolve many of the discrepancies and confirm the actual explosion scenario. Future high resolution and high sensitivity telescopes such as MeerKAT \footnote{http://www.ska.ac.za/science-engineering/meerkat/}, The Square Kilometer Array (SKA) \footnote{http://www.ska.ac.za/} and the Next-Generation VLA \footnote{https://public.nrao.edu/futures/} have the potential to discover many intermediate-age SNe and young ejecta-dominated SNRs in the Local Group and beyond, and provide further observational constraints on the SN-SNR transition phase. Additionally, the historical light curve and the current images of the ejecta of SN 1885A provides a unique consistency check between SN progenitor and SNR evolution models. In a forthcoming paper, we will analyze the latest X-ray observations of SN 1885A by running a full grid of M$_{ch}$ and Sub-M$_{ch}$ models, and a Markov Chain Monte Carlo analysis of the historical light curve and color evolution, and compare with our VLA results.
\acknowledgments

We would like to thank Rob Fesen and Mike Garcia for discussions and suggestions for our work. SKS acknowledges helpful discussions with Ryan Foley and Enrico Ramirez-Ruiz on optical light curves and transients, and Janos Botyanszki and H\'{e}ctor Mart\'{i}nez-Rodr\'{i}guez for guidance with M$_{ch}$ and Sub-M$_{ch}$ models. SKS, LC and CB acknowledge NSF/AST- 1412980 for support of this work. The National Radio Astronomy Observatory is a facility of the National Science Foundation operated under cooperative agreement by Associated Universities, Inc. This research has made use of the NASA Astrophysics Data System
(ADS) Bibliographic Services\\

%

\facilities{VLA, HST}


\software{Numpy \citep{numpy}, Scipy \citep{scipy2001}, Matplotlib \citep{Hunter2007}, Astropy \citep{Astropy2013}, Ipython \citep{Perez2007}, AIPS.}

\bibliography{SN1885_Manuscript}

\end{document}